\documentclass[twocolumn,aps,showpacs,pre]{revtex4}
\usepackage{amsmath}
\usepackage{color}
\usepackage{graphicx}
\usepackage{subfigure}
\usepackage{dcolumn}
\usepackage{bm}
\usepackage{subfloat}
\usepackage[
    colorlinks=true,
    anchorcolor=blue,
    filecolor=green,
    citecolor=blue,
    urlcolor=blue,
    linkcolor=red
 ]{hyperref}

\begin{document}

\title{Crossover from Equilibration to Aging: (Non-equilibrium) Theory vs. Simulations}

\author{P. Mendoza-M\'endez$^{1,\star}$, E. L\'azaro-L\'azaro$^1$, L. E.  S\'anchez-D\'iaz$^{2,\dagger}$,
P. E. Ram\'irez-Gonz\'alez$^{3}$, G. P\'erez-\'Angel$^{4}$, and M. Medina-Noyola$^{1,*}$}

\address{(1)\ Instituto de F\'{\i}sica {\sl ``Manuel Sandoval Vallarta"},
Universidad Aut\'{o}noma de San Luis Potos\'{\i}, \'{A}lvaro
Obreg\'{o}n 64, 78000 San Luis Potos\'{\i}, SLP, M\'{e}xico}

\address{ (2) \ Department of Materials Science and Engineering, University of Tennessee,
Knoxville, Tennessee 37996, USA}

\address{(3) CONACYT-\ Instituto de F\'{\i}sica {\sl ``Manuel
Sandoval Vallarta"}, Universidad Aut\'{o}noma de San Luis Potos\'{\i}, \'{A}lvaro
Obreg\'{o}n 64, 78000 San Luis Potos\'{\i}, SLP, M\'{e}xico}

\address{(4)\ Departamento de F\'isica Aplicada CINVESTAV-IPN, Unidad M\'erida
Apartado Postal 73 Cordemex  97310. M\'erida, Yuc.,  M\'{e}xico}

\date{\today}

\begin{abstract}
Understanding glasses and the glass transition requires
comprehending  the nature of the crossover from the ergodic (or
equilibrium) regime, in which the stationary properties of the
system have no history dependence, to the mysterious glass
transition region, where the measured properties are
non-stationary and depend on the protocol of preparation. In this
work we use non-equilibrium molecular dynamics simulations to test
the main features of the crossover predicted by the
\emph{molecular} version of the recently-developed multicomponent
non-equilibrium self-consistent generalized Langevin equation
(NE-SCGLE) theory. According to this theory, the glass transition
involves the abrupt passage from the ordinary pattern of full
equilibration to the aging scenario characteristic of
glass-forming liquids. The same theory explains that
this abrupt transition will always be observed as a
blurred crossover by the unavoidable finiteness of the time window
of any experimental observation. We find that within their finite
waiting-time window, the simulations confirm the general trends
predicted by the theory.
\end{abstract}


\maketitle

\section{Introduction}\label{section1}
The amorphous solidification of glass- and gel-forming liquids is
an ubiquitous non-equilibrium process of enormous relevance in
physics, chemistry, biology, and materials science and engineering
\cite{dawson}. In contrast with equilibrium crystalline solids,
whose properties have no history dependence, non-equilibrium
amorphous solids may exhibit aging and  their properties actually
depend on their preparation  protocol \cite{angellreview1}.
Although a long and rich theoretical discussion on this subject
has lasted already for several decades, building a general and
fundamental framework that simultaneously predicts the main
universal signatures of these phenomena, as well as their specific
features reflecting the particular molecular interactions and the
concrete fabrication protocol involved, remains ``one of the most
relevant challenges of condensed matter'' \cite{anderson}.

Within the last two decades great advances have been made in the
field of spin glasses, where a mean-field theory has been
developed \cite{cugliandolo} to describe non-equilibrium states.
The models involved, however, cannot describe the evolution of the
spatial structure of real \cite{angellreview1} or simulated
\cite{zaccarelli,hermes,gabriel} \emph{structural} glass formers.
On the other hand, mode coupling theory (MCT) predicts
\cite{goetze1,goetze2} many of the experimentally observed
features of the initial slowdown of real and simulated supercooled
liquids.  As an equilibrium theory, however, it is unable to
describe non-equilibrium phenomena such as aging, and predicts a
divergence of the $\alpha$-relaxation time $\tau_\alpha$ at a
critical temperature $T_c$, which is never observed in practice
\cite{angellreview1,edigerreview1, ngaireview1}.

In recent years, however, a general unifying theory has been
developed, which might well provide the long-awaited fundamental
framework referred to above. This is the \emph{non-equilibrium
self-consistent generalized Langevin equation} (NE-SCGLE) theory
\cite{nescgle1}. This theory was built upon a non-stationary
extension \cite{nescgle1} of Onsager's general and fundamental
laws of linear irreversible thermodynamics and the corresponding
stochastic theory of thermal fluctuations \cite{onsager1,
onsager2,onsagermachlup1,onsagermachlup2}, adequately extended
\cite{delrio,faraday} to allow for the description of memory
effects and spatial non-locality. From this general and abstract
formalism, and after a number of theoretical arguments and
approximations, the concrete but generic NE-SCGLE theory of
irreversible processes in liquids was derived. As summarized
below, this theory simultaneously predicts relevant universal
signatures of the glass and the gel transitions, as well as
specific features reflecting the particular molecular interactions
of the systems considered.

For example, for simple liquids with purely repulsive
interparticle interactions, the NE-SCGLE theory leads to a simple
and intuitive description of the non-stationary and
non-equilibrium process of formation of (high-temperature,
high-density) hard-sphere--like glasses \cite{nescgle3}. For model
liquids with repulsive plus attractive interactions, the NE-SCGLE
theory predicts a still richer and more complex scenario, which
also includes the formation of sponge-like gels and porous glasses
by arrested spinodal decomposition \cite{nescgle5} at low
densities and temperatures. The NE-SCGLE theory has recently been
extended to multi-component systems \cite{nescgle4} and to systems
of non-spherical particles \cite{gory1,nescgle7}, thus opening the
route to the description of more subtle and complex
non-equilibrium amorphous states of matter.

Although these predicted scenarios are qualitatively consistent
with experimental observations, a more critical and quantitative
evaluation is required before this theory can gain acceptance as a
reliable microscopic non-equilibrium statistical thermodynamic
theory. Thus, the main purpose of the present work is to carry out
the first such systematic comparison, using as a reference the
results of the {\emph{molecular dynamics} (MD) simulations of Ref.
\cite{gabriel}, which describe the equilibration and aging of a
\emph{polydisperse}} hard-sphere (HS) liquid. As it will be shown
below, within the time window of the simulations, a remarkable
quantitative agreement is observed between the predicted scenario
and the simulation results.

This paper is structured as follows: The theoretical arguments and
approximations in which the NE-SCGLE theory of irreversible
processes is based are briefly summarize in Section
\ref{section2}. For simplicity, this summary focuses on the
original version of the NE-SCGLE theory which describes the
structure and dynamics of \emph{monocomponent} Brownian liquids.
However, in order to model the polydispersity as well as the
passage from short-time ballistic to long-time diffusive dynamics
involved in the MD simulations, we resort to the {\it molecular}
version of  the recently-developed \emph{multicomponent} NE-SCGLE
theory \cite{nescgle4}. To facilitate the reading of this
manuscript, however, the discussion of these general theoretical
(but rather technical) aspects are collected as Appendices
\ref{appA}-\ref{appD} at the end of the manuscript. Thus, Section
\ref{section3} contains the main results of this work, which
compares the predicted scenario with the simulation results for
the crossover from equilibration to aging of a dense polydisperse
hard-sphere liquid. Finally, the main conclusions and a discussion
of possible directions for further work are contained in Section
\ref{section4}.

\section{Fundamental basis of the NE-SCGLE theory}\label{section2}

As mentioned in the introduction, the non-equilibrium
self-consistent generalized Langevin equation (NE-SCGLE) theory
was derived as a generic application of the non-equilibrium
extension \cite{nescgle1} of Onsager's theory of time-dependent
thermal fluctuations. Here we briefly review the main features of
this abstract and general formalism, and the manner in which it
becomes, in a particular application, a generic theory of the
non-equilibrium evolution of the structure and dynamics of simple
liquids.

\subsection{From a general and abstract formalism to a concrete but generic theory}

For Onsager's theory we mean the general and
fundamental laws of linear irreversible thermodynamics and the
corresponding stochastic theory of thermal fluctuations, as stated
by Onsager \cite{onsager1, onsager2} and by Onsager and Machlup
\cite{onsagermachlup1,onsagermachlup2}, respectively, and as
extended in Refs. \cite{delrio,faraday} to allow for the
description of memory effects and spatial non-locality. The
fundamental assumption of the non-equilibrium extension of
Onsager's theory is that an arbitrary non-equilibrium slow
relaxation process may be described as a globally non-stationary,
but locally stationary, stochastic process \cite{nescgle3}. From
this assumption, general time-evolution equation for the
non-stationary mean value $\overline{a}_i(t_w)$ and covariance
$\sigma_{ij}(t_w)\equiv \overline{\delta a_i (t_w)\delta a _j
(t_w)}$ of the fluctuations $\delta a_i(t_w) =
a_i(t_w)-\overline{a}_i(t_w)$ of the $M$ macroscopic state
variables $[a_1(t_w), a_2(t_w), ..., a_M(t_w)]$ are derived.

To apply this canonical formalism one has to define which physical
properties are represented by the abstract state variables
$a_i(t_w)$. For example, if we have in mind a monocomponent liquid
formed by $N$ particles in a volume $V$, we may identify
$a_i(t_w)$ with the instantaneous number $N_i(t_w)$ of particles
in the volume $\Delta V=V/M$ of the \emph{i}th cell of an
(imaginary) partitioning of the volume $V$ into $M$ cells. Or,
better, with the ratio $n_i(t_w)\equiv N_i(t_w)/\Delta V$, which
in the limit $\Delta V/V \to 0$ becomes the local particle
concentration profile $n(\textbf{r},t_w)$. As explained in detail
in Ref. \cite{nescgle1}, this leads to concrete but generic (i.e.,
applicable to any monocomponent liquid)  time evolution equations
for the mean value $\overline{n}(\textbf{r},t_w)$ and for the
covariance $\sigma(\textbf{r},\textbf{r}';t_w)\equiv
\overline{\delta n (\textbf{r},t_w)\delta n (\textbf{r}',t_w)}$ of
the fluctuation $\delta n (\textbf{r},t_w) = n (\textbf{r},t_w)-
\overline{n}(\textbf{r},t_w)$. The first of these equations reads
\begin{equation} \frac{\partial \overline{n}(\textbf{r},t_w)}{\partial
t_w} = D^0{\nabla} \cdot b(\textbf{r},t_w)\overline{n}(\textbf{r},t_w)
\nabla \beta\mu[{\bf r};\overline{n}(t_w)], \label{difeqdlp}
\end{equation}
whereas the second is written in terms of the Fourier transform
(FT) $\sigma(k;\textbf{r},t_w)$ of the globally non-uniform but
locally homogeneous covariance $\sigma(\textbf{r},\textbf{r} +
\textbf{x};t_w)$,
\begin{eqnarray}
\begin{split}
\frac{\partial \sigma(k;\textbf{r},t_w)}{\partial t_w} = & -2k^2 D^0
\overline{n}(\textbf{r},t_w) b(\textbf{r},t_w)
\mathcal{E}(k;\overline{n}(\textbf{r},t_w)) \sigma(k;\textbf{r},t_w)
\\ & +2k^2 D^0 \overline{n}(\textbf{r},t_w)\ b(\textbf{r},t_w), \label{relsigmadif2p}
\end{split}
\end{eqnarray}
In these equations $D^0$ is the particles' short-time
self-diffusion coefficient \citep{apexD0}, $b(\textbf{r},t_w)$ is their local
reduced mobility, and $\mu[{\bf r};\overline{n}(t_w)]$ is their
chemical potential. $\mathcal{E}(k;\overline{n}(\textbf{r},t_w))$
is the FT of $\mathcal{E}[{\bf r},\textbf{r} + \textbf{x};n]\equiv
\left[ {\delta \beta\mu [{\bf r};n]}/{\delta n(\textbf{r} +
\textbf{x})}\right]$.

Eqs. (\ref{difeqdlp}) and (\ref{relsigmadif2p}) above correspond
to Eqs. (4.1) and (4.3) of Ref. \cite{nescgle1}, which discusses
other more specific theories and limits that turn out to be
contained as particular cases of these equations. For example, let
us imagine that we manipulate the system to an arbitrary
(generally non-equilibrium) initial state with mean concentration
profile $\overline{n}^{0}(\textbf{r})$ and covariance
$\sigma^{0}(k;\textbf{r})$, for then letting the system
equilibrate for $t_w>0$ in the presence of an external field
$\psi({\bf r})$ and in contact with a temperature bath of
temperature $T$. The solution of Eqs. (\ref{difeqdlp}) and
(\ref{relsigmadif2p})  then describes how the system relaxes to
its final equilibrium state whose mean profile and covariance are
$\overline{n}^{eq}(\textbf{r})$ and $\sigma^{eq}(k;\textbf{r})$.
Describing this response at the level of the mean local
concentration profile $\overline{n}(\textbf{r},t_w)$ is precisely
the aim of \emph{dynamic} density functional theory (DDFT)
\cite{tarazona1, tarazona2, archer}, whose central equation is
recovered from Eq. (\ref{difeqdlp}) in the limit in which we
neglect the friction effects embodied in $b(\textbf{r},t_w)$ by
setting $b(\textbf{r},t_w)=1$ (see Eq. (15) of Ref.
\cite{tarazona1}).

The description of the non-equilibrium state of the system in
terms of the random variable $n(\textbf{r},t_w)$ is not complete,
however, without the simultaneous description of the relaxation of
the covariance $\sigma(k;\textbf{r},t_w)$ in Eq.
(\ref{relsigmadif2p}). In fact, under some circumstances, the main
signature of the non-equilibrium evolution of a system may be
embodied not in the temporal evolution of the mean value
$\overline{n}({\bf r};t_w)$ but in the evolution of the covariance
$\sigma(k;\textbf{r},t_w)$ (which is essentially a non-uniform and
non-equilibrium version of the static structure factor). This may
be the case, for example, when a homogeneous system in the absence
of external fields remains approximately homogeneous,
$\overline{n}({\bf r};t_w)\approx \overline{n}\equiv N/V$, after a
sudden temperature change. Under these conditions, the
non-equilibrium process is described only by the solution of Eq.
(\ref{relsigmadif2p}). Let us point out that in the limit $b(t_w)
\to 1$ and within the small-wave-vector approximation,
$\mathcal{E}(k;\overline{n})\approx \mathcal{E}_0 +\mathcal{E}_2
k^2$,  Eq. (\ref{relsigmadif2p}) becomes the basic kinetic
equation describing the early stage of spinodal decomposition
(see, for example, Eq. (3.4) of Ref. \cite{furukawa}).

Eqs. (\ref{difeqdlp}) and (\ref{relsigmadif2p}) above are coupled
between them through the local mobility function
$b(\textbf{r},t_w)$, essentially a non-stationary and
state-dependent Onsager's kinetic coefficient. In addition, these
two equations are also coupled, through  $b(\textbf{r},t_w)$, with
the two-point (van Hove) correlation function $C(\textbf{r},\tau;
\textbf{x};t_w)\equiv \overline{\delta n(\textbf{x},t_w)\delta n
(\textbf{x}+\textbf{r},t_w+\tau)}$. According to Ref.
\cite{nescgle1}, the memory function of $C(\textbf{r},\tau;
\textbf{x};t_w)$ can in its turn be  written approximately in
terms of $\overline{n}({\bf r};t_w)$ and
$\sigma(k;\textbf{r},t_w)$, thus introducing strong non-linear
effects. Thus, even before solving Eqs. (\ref{difeqdlp}) and
(\ref{relsigmadif2p}), they reveal a number of relevant features
of general and/or universal character.

The most illuminating of them is that, besides the equilibrium
stationary solutions $\overline{n}^{eq}(\textbf{r})$ and
$\sigma^{eq} (k;\textbf{r})$, defined by the equilibrium
conditions $\nabla \beta\mu[{\bf r};\overline{n}^{eq}]=0$ and
$\mathcal{E}(k;\overline{n}(\textbf{r},t_w))
\sigma(k;\textbf{r},t_w)=1$, Eqs. (\ref{difeqdlp}) and
(\ref{relsigmadif2p}) also predict the existence of another set of
stationary solutions that satisfy the dynamic arrest condition,
$\lim_{t_w\to \infty} b(\textbf{r},t_w)=0$. This far less-studied
second set of solutions describes, however, important
non-equilibrium stationary states of matter, corresponding to
common and ubiquitous non-equilibrium amorphous solids, such as
glasses and gels.

\subsection{Spatial uniformity, a simplifying approximation.}

To appreciate the essential physics of this fundamental and
universal prediction of Eqs. (\ref{difeqdlp}) and
(\ref{relsigmadif2p}), the best is to provide explicit examples.
To do this without a high mathematical cost, however, let us write
$\overline{n}(\textbf{r},t_w)$ as $\overline{n}(\textbf{r},t_w)=
\overline{n}(t_w) + \Delta \overline{n}(\textbf{r},t_w)$, and in a
first stage let us neglect the spatial heterogeneities represented
by the deviations $\Delta \overline{n}(\textbf{r},t_w)$. As a
result, rather than solving the time-evolution equation for
$\overline{n}({\bf r};t_w)$, we have that $\overline{n}(t_w)$ now
becomes a control parameter, so that we only have to solve the
time-evolution equation for the covariance
$\sigma(k,\textbf{r};t_w)$. We may consider, for example, the
specific case in which the system is {\it constrained} to remain
isochoric and spatially {\it homogeneous} ($\overline{n}({\bf
r};t_w)\approx \overline{n}\equiv N/V$) after an instantaneous
temperature quench at time $t_w=0$, from an arbitrary initial
temperature to a lower final temperature $T$. For this process,
the time-evolution equation for the Fourier transform (FT)
$\sigma(k;t_w)$ of the covariance $\sigma(\textbf{r},
\textbf{r}';t_w)=\sigma(\mid\textbf{r}-\textbf{r}'\mid;t_w)$ can
be written, for $t_w>0$ and in terms of the non-stationary static
structure factor $S(k;t_w)\equiv \sigma(k;t_w)/\overline{n}$, as

\begin{equation}
\frac{\partial S(k;t_w)}{\partial t_w} = -2k^2 D^0
b(t_w)\overline{n}\mathcal{E}_f(k) \left[S(k;t_w)
-1/\overline{n}\mathcal{E}_f(k)\right]. \label{relsigmadif2pp}
\end{equation}
in which $\mathcal{E}_f(k)=\mathcal{E}(k;\overline{n},T_f)$ is the Fourier
transform (FT) of the functional derivative
$\mathcal{E}[\mid\textbf{r}-\textbf{r}'\mid;n,T] \equiv \left[
{\delta \beta\mu [{\bf r};n]}/{\delta n({\bf r}')}\right]$ of the
chemical potential $\mu$, evaluated at $n({\bf r})=\overline{n}$
and $T=T_f$.

It is important to mention that the solution of this
equation yields in principle $S(k;t_w)$ as output, for given
$b(t_w)$ provided as input. This calls for an independent
relationship between these two unknowns, which may have the format
of an equation (or system of equations) that accepts $S(k;t_w)$ as
input and yields $b(t_w)$ as output. This is precisely the role of
the following set of equations. The first of them is an expression
for the time-evolving mobility $b(t_w)$,
\begin{equation}
b(t_w)= [1+\int_0^{\infty} d\tau\Delta{\zeta}^*(\tau; t_w)]^{-1},
\label{bdt}
\end{equation}
in terms of the $t_w$-evolving, $\tau$-dependent friction
coefficient $\Delta{\zeta}^*(\tau; t_w)$, which can be
approximated by \cite{nescgle1}
\begin{equation}
\begin{split}
  \Delta \zeta^* (\tau; t_w)= \frac{D_0}{24 \pi
^{3}\overline{n}}
 \int d {\bf k}\ k^2 \left[\frac{ S(k;
t_w)-1}{S(k; t_w)}\right]^2 \times \\
F(k,\tau; t_w)F_S(k,\tau; t_w).
\end{split}
\label{dzdtquench}
\end{equation}

In this equation $\tau$ is the \emph{correlation time} and $t_w$
is the \emph{waiting} (or evolution) time. $F(k,\tau; t_w)$ and
$F_S(k,\tau; t_w)$ are, respectively, the collective and self
non-equilibrium intermediate scattering functions (ISFs), whose
respective memory functions are approximated to yield the
following approximate expressions for the Laplace transforms (LT)
$\hat F(k,z; t_w)$ and $\hat F_S(k,z; t_w)$,
\begin{gather}\label{fluctquench}
\hat  F(k,z; t_w) = \frac{S(k; t_w)}{z+\frac{k^2D^0 S^{-1}(k;
t_w)}{1+\lambda (k;t_w)\ \Delta \hat  \zeta^*(z; t_w)}},
\end{gather}
and
\begin{gather}\label{fluctsquench}
\hat  F_S(k,z; t_w) = \frac{1}{z+\frac{k^2D^0 }{1+\lambda (k;t_w)\
\Delta \hat \zeta^*(z; t_w)}}.
\end{gather}
In these equations $\lambda (k)$ is a phenomenological
``interpolating function" \cite{nescgle1}, given by
\begin{equation}
\lambda (k;t_w)=1/[1+( k/k_{c}(t_w)) ^{2}], \label{lambdadk}
\end{equation}
with $k_c(t_w)$ being an empirically chosen cutoff wave vector.

Eqs. (\ref{dzdtquench})-(\ref{lambdadk}) are the non-equilibrium
extension of the corresponding equations of the equilibrium SCGLE
theory, which is recovered in the long-$t_w$ stationary limit in
which $S(k;t_w\to\infty) \to S^{(eq)}(k) \equiv
1/\overline{n}\mathcal{E}_f(k)$. The derivation of these equations
in Ref. \cite{nescgle1} also extends to non-equilibrium
conditions the same approximations and assumptions employed in the
original derivation of the equilibrium SCGLE theory \cite{todos2}.
Such an extension is quite natural within the framework of the
non-equilibrium generalization of Onsager's theory, but not in the
context of the Mori-Zwanzig  formalism \cite{boonyip},  which is deeply rooted in
the equilibrium condition.

Coupling Eqs. (\ref{relsigmadif2pp}) and (\ref{bdt}) with Eqs.
(\ref{dzdtquench})-(\ref{lambdadk}) results in  the NE-SCGLE
closed system of equations that must be solved self-consistently.
Thus, the simultaneous solution of Eqs.
(\ref{relsigmadif2pp})-(\ref{lambdadk}) above constitutes the
NE-SCGLE description of the spontaneous evolution of the structure
and dynamics of an \emph{instantaneously} and \emph{homogeneously}
quenched \emph{monocomponent} liquid. The only element that we
still have to determine is the empirically chosen cutoff wave
vector $k_c(t_w)$. For simplicity, we shall define this parameter
in reference to the position of the main peak of  $S(k;t_w)$, in
an identical manner as the cutoff wave vector $k_c^{eq}$ of the
equilibrium SCGLE theory is defined in reference to the position
of the main peak of  $S^{(eq)}(k)$. In this manner, the NE-SCGLE
theory becomes a self-consistent theory with no adjustable
parameters.

\subsection{General physical insights revealed by the NE-SCGLE equations}

Being a particular case of Eq. (\ref{relsigmadif2pp}), the most
relevant and general physical insights provided by Eq.
(\ref{relsigmadif2pp}) is the NE-SCGLE prediction of the existence
of two fundamentally different kinds of stationary solutions,
implying the existence of two fundamentally different kinds of
states of matter. The first corresponds to ordinary thermodynamic
equilibrium states, in which stationarity is attained because the
factor $\left[S(k;t_w) -1/\overline{n}\mathcal{E}_f(k)\right]$ on
the right side of Eq. (\ref{relsigmadif2pp}) vanishes, i.e.,
because $S(k;t_w)$ is able to reach its thermodynamic equilibrium
value $S^{(eq)}(k;\overline{n},T_f)=
1/\overline{n}\mathcal{E}_f(k)$, while the mobility $b(t_w)$
attains a \emph{finite} positive long-time limit $b_f$.

Under these conditions, one can estimate the equilibration time $t^{eq}_w(\overline{n},T_f)$ of a quench to a
final temperature $T_f$ at fixed density $\overline{n}$ (or fixed volume fraction $\phi$),
as the waiting time such that the difference between $S(k_{max};t_w)$ and its asymptotic
equilibrium value  $S^{(eq)}(k_{max};\overline{n},T_f)$ is sufficiently small,
say $[S(k_{max};t_w)-S_f(k_{max})]\approx e^{-5}$. Thus, according to the solution of Eq. (\ref{relsigmadif2pp}),
the condition defining $t^{eq}_w(\phi)$ is $[S(k_{max};t^{eq}_w)-S_f(k_{max})]= \exp [-2k^2D_0 u(t^{eq}_w)/S_f(k_{max})]\approx \exp[- 5]$, where $u(t_w)\equiv \int_0^{t_w} b(t_w') dt_w'$. Since for long waiting times $u(t_w)\propto b_f t_w$, later in this paper we shall estimate $t^{eq}_w(\phi)$ as

\begin{equation}
t_w^{eq}(\overline{n},T_f) \approx 5
S^{(eq)}(k_{max};\overline{n},T_f) /2k_{max}^2D^0b_f,
\label{tequil}
\end{equation}
where $D^0b_f=D^{(eq)}_L(\overline{n},T_f)$ is the
\emph{equilibrium} long-time self-diffusion coefficient at the
final state point $(\overline{n},T_f)$. This equilibration time is
predicted to increase when $b_f$ decreases, and to diverges as
$1/b_f$   when the state point $(\overline{n},T_f)$ approaches the
ergodic--non-ergodic transition line. This means that already in
the ergodic neighborhood of this boundary one should experience
enormous difficulties in equilibrating the system within practical
experimental times.

The second class of stationary solutions of Eq.
(\ref{relsigmadif2pp}) emerges from the possibility that the
long-time asymptotic limit of the kinetic factor $b(t_w)$
vanishes, so that $dS(k;t_w)/dt_w$ vanishes at long times without
requiring the equilibrium condition $\left[S(k;t_w)
-1/\overline{n}\mathcal{E}_f(k)\right]=0$ to be fulfilled. Under
these conditions $S(k;t_w)$ will now approach a distinct
non-equilibrium stationary limit, denoted by $S_a(k)$, which is
definitely different from the expected equilibrium value $S_f(k)=
S^{(eq)}(k;\overline{n},T_f)$. Furthermore, the difference
$[S(k;t_w)-S_a(k)]$ is predicted to decay to zero in an extremely
slow fashion, namely, as $t_w^{-0.833}$ \cite{nescgle3}. This
second class of stationary solutions represents dynamically
arrested states of matter (glasses, gels, etc.). The properties of
these stationary but intrinsically non-equilibrium states, such as
$S_a(k)$, are predicted to strongly dependent on the preparation
protocol (in our example, on $T_i$ and $T_f$). Furthermore, due to
the extremely slow approach to its asymptotic limit, no matter how
long we wait, any finite-time measurement will only record the
non-stationary, $t_w$-dependent value of the measured properties
($S(k;t_w)$, $F(k,\tau; t_w)$, $F_S(k,\tau; t_w)$, etc.).

Although the NE-SCGLE system of equations
(\ref{dzdtquench})-(\ref{lambdadk}) is highly non-linear, changing
variable from $t_w$ to $u(t_w)\equiv \int_0^{t_w} b(t'_w)dt'_w$
re-writes Eq. (\ref{relsigmadif2pp}) as a linear relaxation
equation for $S^*(k;u)$,
\begin{equation}
\frac{\partial }{\partial u}\left[S^*(k;u) -S_f(k)\right] =
-\alpha (k) \left[S^*(k;u) -S_f(k)\right],
\label{relsigmadif2pp2}
\end{equation}
with $\alpha (k) \equiv 2k^2D^0/S_f(k)$. The solution of Eq.
(\ref{relsigmadif2pp}) can thus be written as
$S(k;t)=S^*(k;u(t))$, with
\begin{equation} S^*(k;u)=S_f(k)+ \left[S_i(k)-S_f(k)\right]e^{-\alpha
(k)u}. \label{solsdktexp}
\end{equation}

It also predicts \cite{nescgle3} that the non-linearity is
actually encapsulated in the time-dependence of the ``internal''
(or ``material'') time $u(t_w)$, in full consistency with the
phenomenological model of aging of Tool and Narayanaswamy
\cite{tool,narayanaswamy}, commonly used to model aging  and to
fit a large number of experimental data
\cite{hornboll,hecksher,richert}. We thus conclude that the
NE-SCGLE theory captures this intriguing and relevant universality
and casts it in a more fundamental and precise first-principles
physical context.

\section{Crossover from ergodic equilibration to non-equilibrium aging
of a polydisperse hard-sphere liquid}\label{section3}

In this section we discuss the quantitative test of a third general insight of the NE-SCGLE theory. This
refers to the nature of the high density hard-sphere glass transition. According to the scenario
predicted by the  NE-SCGLE theory, the discontinuous and singular transition predicted by equilibrium
theories (such as MCT or the equilibrium SCGLE theory) for the hard-sphere liquid is intrinsically
correct, but essentially unobservable in practice. This is due to the fact that such theories predict
the divergence of the \emph{equilibrium} $\alpha$-relaxation time $\tau^{(eq)}_\alpha(\phi)$ at the
critical volume fraction  $\phi_a$ (and that it remains infinite for $\phi\ge \phi_a$). Of course,
if $\tau^{(eq)}_\alpha(\phi)$ becomes infinite, it is reasonable to conjecture that also
the \emph{equilibration} time $t^{(eq)}(\phi)$ (i.e., the time it takes the system to equilibrate
after preparation) must also be infinite. If this conjecture were correct, then the predicted
diverging equilibrium scenario will not be amenable to experimental tests, due to the unavoidable
constraint of any real experiment or measurement, to be limited to finite time-windows.

Let us mention that the previous scenario, in which the control parameter is the volume fraction $\phi$, is also expected to hold almost without change when we consider a sequence of quenches from a common initial temperature $T_0$ to a final temperature $T$ along the same isochore. In this case, the control parameter is the temperature $T$, with its inverse $1/T$ playing the role of the volume fraction $\phi$ in the present discussion. This $\phi \leftrightarrow  1/T$ correspondence has been predicted by the equilibrium SCGLE theory (see ref. \cite{dtsoft}) and by
the present non-equilibrium extension (separate manuscript).
It is a fact, however, that in any real experiment (or simulation) one indeed
determines a ``real'' experimental value $\tau_\alpha(\phi;t_w)$ (or $\tau_\alpha(1/T;t_w)$) of
the the $\alpha$-relaxation time. In general, however, such measured value $\tau_\alpha(t_w)$ will
depend on the waiting time $t_w$ after preparation, thus being a non-equilibrium property that
cannot be predicted by an equilibrium theory. The power of the NE-SCGLE theory is precisely that
it provides a detailed prediction of the non-equilibrium evolution of the system at any finite
evolution time $t_w$, thus shifting the attention from unobservable infinite-time equilibrium
singularities, to the finite-$t_w$ non-equilibrium properties actually measured in practice, such as $\tau_\alpha(t_w)$

\begin{figure}[h]
\centering
\subfigure{ \label{fig:1a}
\includegraphics[scale=.3]{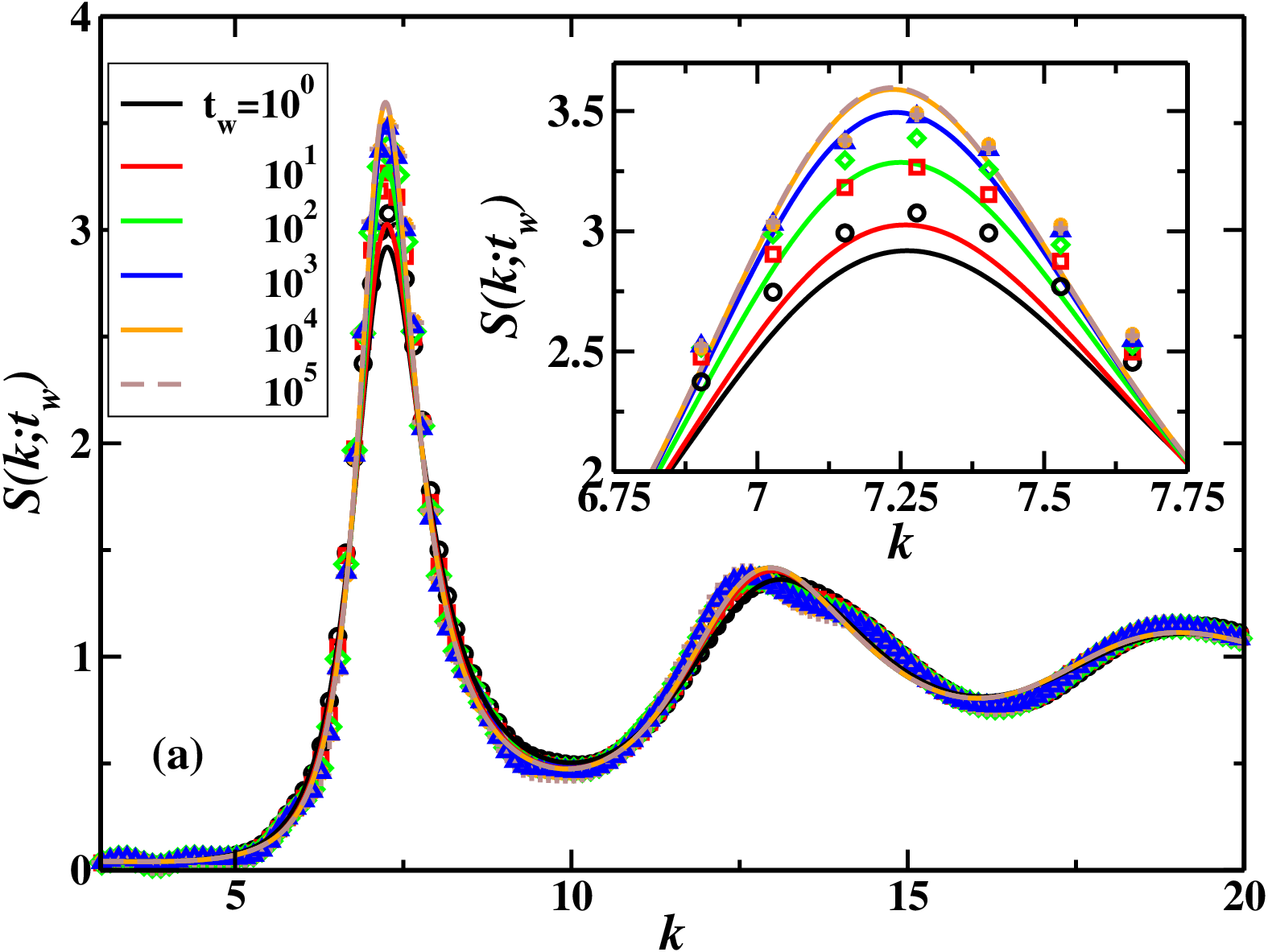}
}
\subfigure{ \label{fig:1b}
\includegraphics[scale=.3]{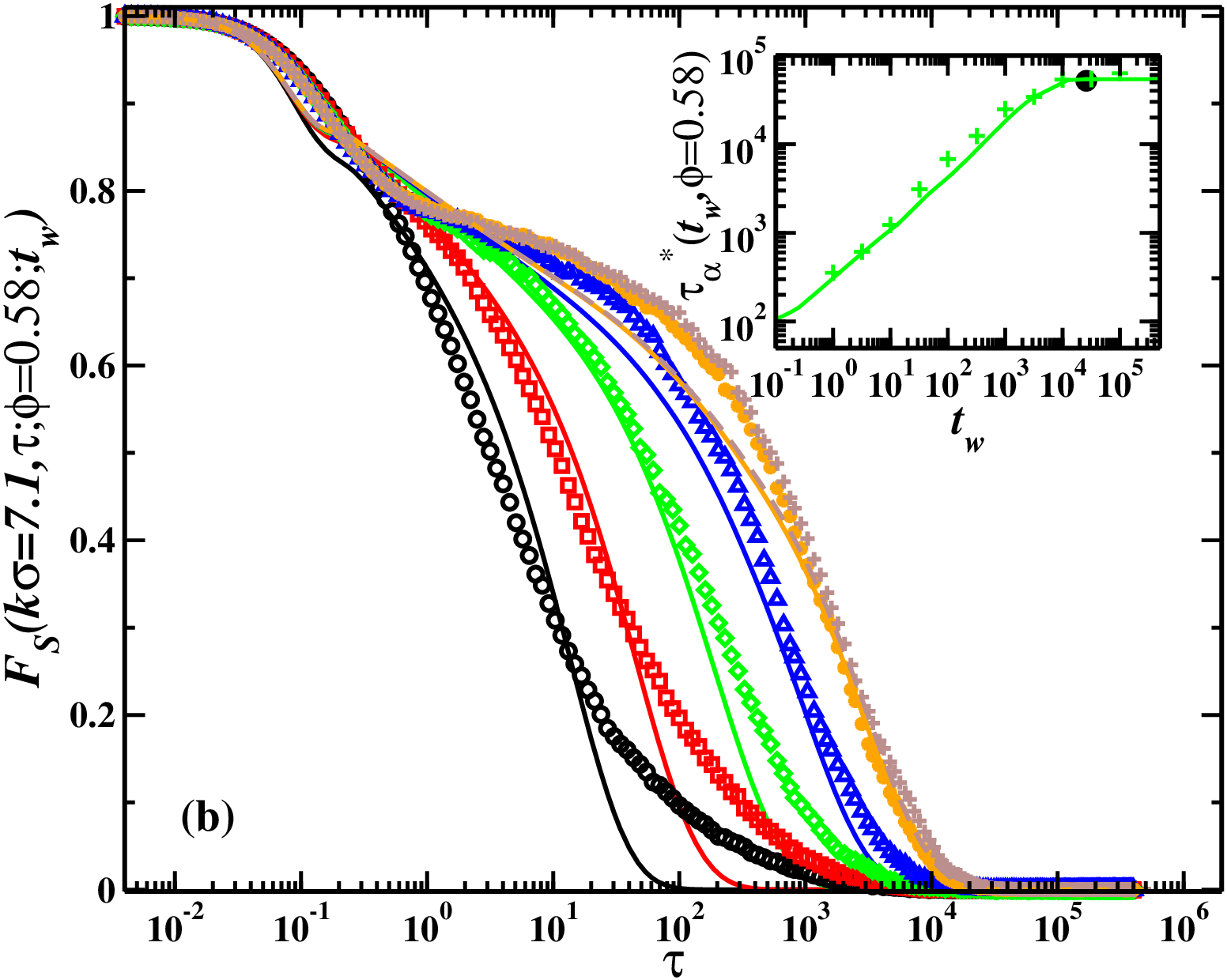}
}
\caption{Non-equilibrium evolution of the  structure and dynamics
of a hard-sphere liquid in the process of its isochoric
equilibration at fixed volume fraction  $\phi=0{.}58$ according to
the (MD) non-equilibrium simulations (symbols) and to the NE-SCGLE
theory (solid lines). Fig. \ref{fig:1a} Snapshots of $S(k,t_w)$ as a function of
$k$, with a zoom at the main peak in the inset, for the indicated
sequence of waiting times. Fig. \ref{fig:1b} Corresponding snapshots (from left
to right) of $F_S(k,\tau;\phi,t_w)$ plotted as a function of
correlation time $\tau$ at fixed $k\sigma=7{.}1$.  The inset plots
the $\alpha$-relaxation time $\tau_{\alpha}(t_w,\phi)$, scaled as
$\tau_{\alpha}^*(t_w,\phi)\equiv k^2D^0\tau_{\alpha}(k\sigma;t_w,\phi)$, as
a function of evolution time $t_w$, for this equilibration process. The dark
asterisk is the equilibration point
($t_w^{eq},\tau_\alpha^{*eq}$). } \label{fig1}
\end{figure}

These are precisely the predictions that we mean to quantitatively
test with the following comparisons. In the particular simulations
of Ref. \cite{gabriel}, a hard-sphere liquid was driven to a
non-equilibrium state by means of an effective sudden compression
protocol to a final density $\overline{n}$ corresponding to the
desired volume fraction $\phi=\pi \overline{n}\sigma^3/6$. This
protocol is used to generate an ensemble of configurations
representative of such non-equilibrium state, characterized by a
well-defined initial static structure factor $S_0(k;\phi)$. These
representative configurations are then taken as the initial
condition of an ensemble of standard MD simulation runs describing
the non-equilibrium structural relaxation leading to the
equilibration (or aging) of the system. This non-equilibrium
transient is the subject of study of these simulations (in
contrast with ordinary equilibrium simulations, in which this
stage is discarded).

The theoretical modeling of the same transient is provided by the
simultaneous solution of Eqs.
(\ref{relsigmadif2pp})-(\ref{lambdadk}) above, after complementing
Eq. (\ref{relsigmadif2pp}) with the initial condition
$S(k;t=0)=S_0(k;\phi)$ and after determining the thermodynamic
function $\mathcal{E}_f(k)=\mathcal{E}(k;\overline{n},T_f)$
evaluated at the final state point of the quench. In the present
case this corresponds to setting
$\overline{n}\mathcal{E}_f(k)=\overline{n}\mathcal{E}_{HS}(k;\phi)=1/S^{(eq)}_{HS}(k;\phi)$,
for which we use Percus-Yevick's approximation \cite{percusyevick}
with its Verlet-Weis correction  \cite{verletweis}. For the
initial non-equilibrium structure factor $S_0(k;\phi)$ we could
use directly the result of the simulated non-equilibrium
preparation protocol described in the previous paragraph.

Alternatively, we could theoretically model this
\emph{non-equilibrium} structure factor by the \emph{equilibrium}
structure factor of the hard-sphere liquid,
$S^{(eq)}_{HS}(k;\phi_i)$, at an ``initial'' volume fraction
$\phi_i$, chosen such that the structural and/or dynamical
properties of such equilibrium HS liquid are similar to those of
the non-equilibrium state generated by the actual non-equilibrium
preparation protocol. In fact, due to  dynamical equivalence
between soft- and hard-sphere liquids, we could model
$S_0(k;\phi)$ by the equilibrium static structure factor
$S^{(eq)}(k;n_i,T_i)$ of \emph{any} soft-sphere liquid included in
the hard-sphere dynamic universality class \cite{dtsoft,atomic3},
provided that the density $n_i$ and temperature $T_i$ are chosen
such that the structural and/or dynamical properties match those
of the previously defined hard-sphere liquid,
$S^{(eq)}(k;n_i,T_i)\approx S^{(eq)}_{HS}(k;\phi_i)$.

In practice, however, the scenario predicted by the solution of
Eqs. (\ref{relsigmadif2pp})-(\ref{lambdadk}) is virtually
independent of the specific manner to model the initial
non-equilibrium structure factor $S_0(k;\phi)$. Thus, in the
results that follow, we approximated $S_0(k;\phi)$ by the
equilibrium static structure factor  $S^{(eq)}(k;\phi_i,T_i)$ of a
polydisperse fluid of soft spheres of diameter $\overline{\sigma}$
and whose interactions are modeled by the Weeks-Chandler-Andersen
(WCA) pair potential. In this way, the process start with the
system initially at a fluid-like state of temperature
$T_i=0.06[\epsilon/k_B]$ and the same volume fraction $\phi$ of
the simulated HS liquid and, at $t_w=0$, the temperature is
instantaneously lowered to a final value $T_f=0$ at which the
expected equilibrium state is that of a polydisperse hard-sphere
liquid at volume fraction $\phi$.

Fig. \ref{fig1} illustrates the simplest and most straightforward
comparison between the NE-SCGLE theoretical predictions and the
simulation results for the  non-equilibrium isochoric evolution at
fixed $\phi=0.58$ of the HS liquid, in terms of $S(k;t_w)$ and of
the non-equilibrium self intermediate scattering function
$F^{S}(k,\tau;t_w)$ ($\equiv \langle \exp [i\textbf{k}\cdot \Delta
\textbf{R}(t_w)] \rangle$, with $\Delta
\textbf{R}(t_w)=\textbf{R}(t_w+\tau)-\textbf{R}(t_w)$ being the
displacement of a tagged particle). This comparison involves a
sequence of snapshots of $ S(k;t_w)$ as a function of $k$ (Fig.
\ref{fig:1a}), and of $F^{S}(k=7.1\sigma^{-1},\tau;t_w)$ as a
function of correlation time $\tau$ (Fig. \ref{fig:1b}),
corresponding to  a sequence of waiting times $t_w= \ 10^0,\
10^1,\ 10^2,\ 10^3,\ 10^4$ and $10^5$ (in molecular time units,
$[\sigma\sqrt{M/k_BT}]$).

These results illustrate that both, simulations and theory, agree
in that no dramatic changes are observed in the evolution of the
structure, except for the modest increase in the main peak of
$S(k;t_w)$, zoomed-in in the inset of Fig. \ref{fig:1a}. In
contrast, the dynamics does exhibit a remarkable slowing down,
occurring within an ``equilibration'' time $t_w^{eq}(\phi)$. The
kinetics of this equilibration process is best summarized by the
$t_w$-dependence of the non-equilibrium $\alpha$-relaxation time
$\tau_\alpha(t_w,\phi)$, defined here by the condition
$F_S(k=7.1\sigma^{-1},\tau_{\alpha};t_w,\phi) = 1/e$, and
illustrated in the inset of Fig. \ref{fig:1b} for the
equilibration process of the HS liquid at $\phi=0.58$.

\begin{figure}[h]
\includegraphics[scale=.3]{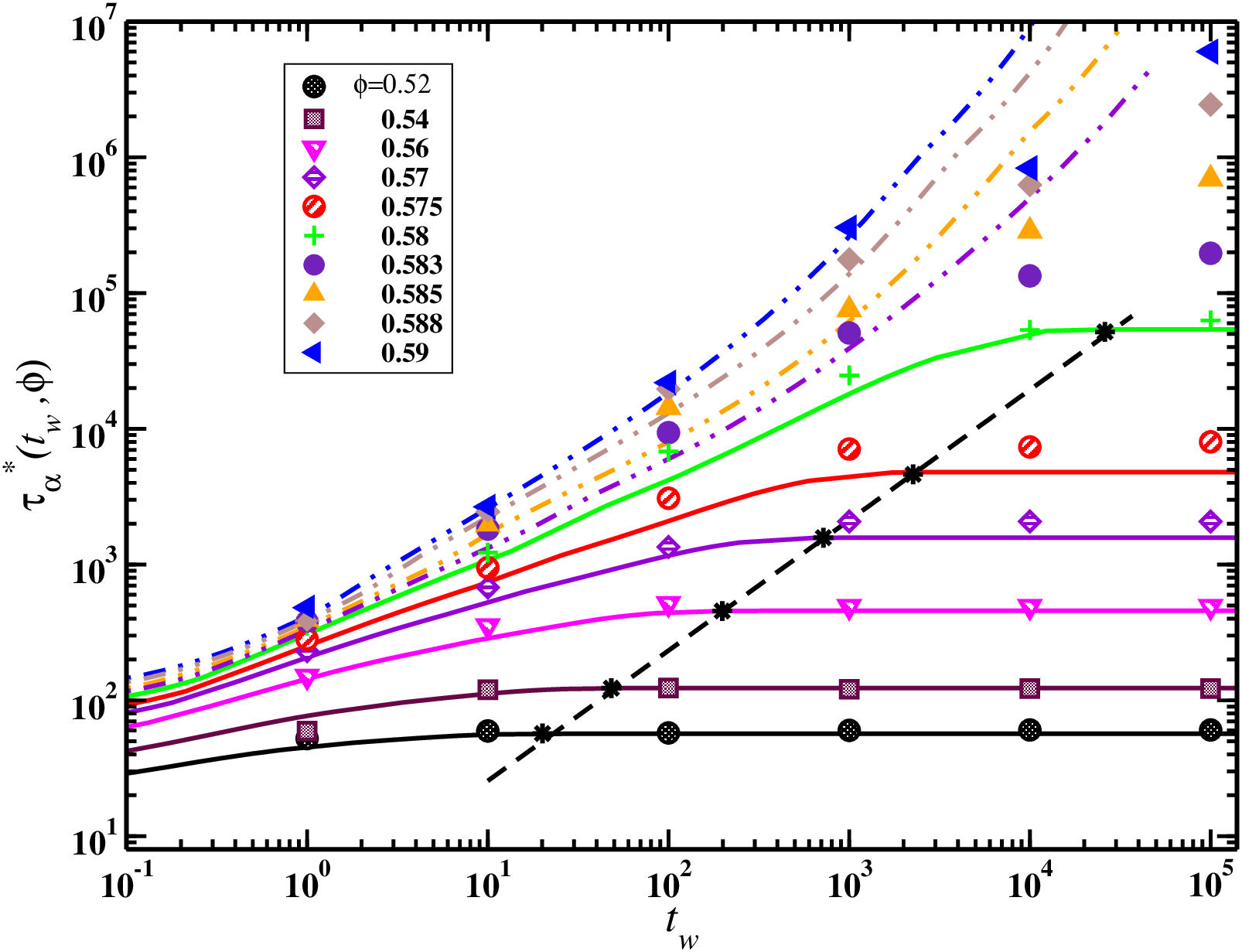}
\caption{Non-equilibrium Molecular Dynamics simulations (symbols)
and NE-SCGLE theoretical results (solid lines for equilibration and dot-dashed lines for aging processes) for the $\alpha$-relaxation time $\tau_{\alpha}^*(t_w,\phi)$, plotted
as a function of evolution time $t_w$ for a sequence of fixed
volume fractions. The asterisks represent the equilibration points  $(t^{eq}_w(\phi), \tau^{*eq}_{\alpha}(\phi))$ and the dashed line passing through them is the power-law fit $\tau_\alpha^{*eq}(\phi)=
2.8\times [t_w^{eq}(\phi)]^{0.96}$.} \label{fig2}
\end{figure}

At this point let us recall that, in order to prevent
crystallization, the MD simulations in the figures actually
correspond to an 8.66\% (size) \emph{polydisperse} HS liquid. To
properly take this fact into account,  the solid lines in Fig.
\ref{fig1}  actually correspond to the solution of the NE-SCGLE
equations for a polydisperse HS liquid, modeled as an equimolar
binary mixture with size ratio yielding a polydispersity of
8.66\%. Similarly, to properly compare the NE-SCGLE theoretical
predictions (originally derived for Brownian, rather than
molecular, liquids) with the present MD simulation data, we
applied the long-time dynamic equivalence between Brownian and
molecular systems proposed in Ref. \cite{AtomicMix}, to adapt the
NE-SCGLE theory to liquids with underlying molecular microscopic
dynamics. This allows us to  compare  on an equal footing the
theoretically-predicted results with the simulated dynamics of the
atomic liquid. These methodological aspects of our theoretical
calculations are explained in Appendices \ref{appA}-\ref{appC}.

By extending the calculations and comparisons in Fig. \ref{fig:1b}
to a sequence of other volume fractions in the metastable region of
the HS liquid, a more panoramic view emerges of the consistency between
the scenarios revealed by theory and by simulations. The results are
presented in Fig. \ref{fig2}, which illustrates the extent of the
consistency between the main qualitative features of the predicted and
the simulated scenarios. For example, in both we see that when the
fixed volume fraction is smaller than 0.582, the system will equilibrate
within a $\phi$-dependent equilibration time $t_w^{eq}(\phi)$ determined by Eq.
(\ref{tequil}). This equilibration time strongly increases with $\phi$, in a very similar
manner as the equilibrium value $\tau_\alpha^{eq}(\phi)$ of the $\alpha$-relaxation
time. In fact, as can be gathered from the asterisks in the figure, our theory
predicts that $t_w^{eq}(\phi)\propto [\tau_\alpha^{eq}(\phi)]^\eta$,
with $\eta \approx 1$ (rather than $\eta \approx 1.43$, as determined in
the simulations \cite{gabriel,kimsaito}).

For $\phi \ge 0.582$, the NE-SCGLE theory agrees with its
equilibrium version (and with MCT) in the prediction that
$\tau_\alpha^{eq}(\phi)$, and hence also $t_w^{eq}(\phi)$, is
infinite. This prediction cannot be refuted nor demonstrated,
since in practice one can only measure \emph{finite}
$\tau^*_\alpha(t_w,\phi)$ at \emph{finite} waiting times, within
finite correlation-time windows. Such finite measurements,
however, constitute a stringent and valuable test of the NE-SCGLE
theory, which always predicts a finite value for
$\tau^*_\alpha(t_w,\phi)$ at any finite $t_w$. The result of such
test is illustrated in Fig. \ref{fig2} with the four irreversible
processes occurring at fixed volume fractions in the non-ergodic
regime $\phi \ge 0.582$ (indicated with fill symbols).

For these processes we observe excellent quantitative agreement
with the simulation data for $t_w \le 10^3$ , but noticeable
deviations at longer $t_w$. The origin of these deviations might lie in the
intrinsic inaccuracies of the approximations involved in the
NE-SCGLE theory and/or in the difficulties to simulate the
relaxation of a genuine non-ergodic system. For example, for
simplicity our theory approximates the mean local density
$\overline n (\textbf{r};t_w)$ by its bulk value $n$, thus
neglecting structural and dynamical heterogeneities. From the
simulation side, the non-equilibrium ensemble employed (see
details in the appendix \ref{appD}) involved at least 40
realizations and $1024$ particles. Although this is perfectly
adequate for a conventional equilibration process, it is perhaps
insufficient at long waiting times in the true non-ergodic regime,
as $\phi$ increases far above $\phi_c \approx 0.582$.

\begin{figure}[h]
\includegraphics[scale=.3]{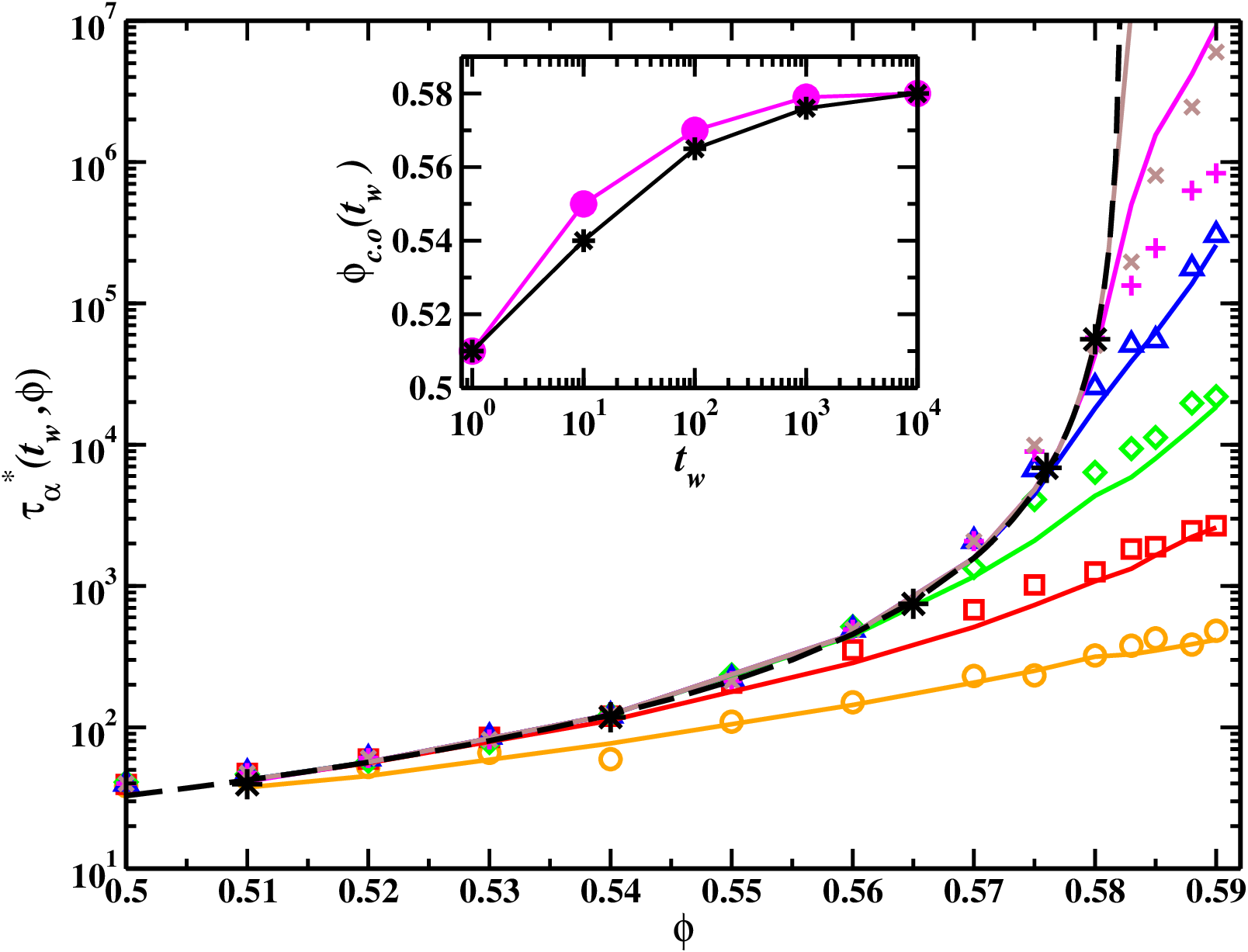}
\caption{Same as in Fig. \ref{fig2}, but now $\tau_{\alpha}^*(t_w,\phi)$ is
plotted as a function of $\phi$ for fixed times $t_w=10^0,\ 10^1,\
10^2,\ 10^3,$ $10^4$ and $10^5$ (from bottom to top; the dashed
line corresponds to $t_w=\infty$). In this case, the dark
asterisks indicate the (theoretical) $t_w$-dependent volume
fraction $\phi_{c.o.}(t_w)$ that describes the crossover from
fully equilibrated to insufficiently equilibrated conditions. In
the inset we compare the predicted (asterisks) and simulated
(circles) results for $\phi_{c.o.}(t_w)$.} \label{fig3}
\end{figure}

Although these limitations of the theory and of the simulations
must be the subject of more detailed and systematic study, the
comparison in Fig. \ref{fig2} is already highly instructive and
revealing, since it provides a kinetic conceptual framework to
discuss the nature of the  processes of equilibration and aging.
To illustrate this,  let us now plot the $\alpha$-relaxation time
$\tau^*_{\alpha}(t_w,\phi)$ as a function of $\phi$ for a sequence
of fixed waiting times. Fig. \ref{fig3} illustrates that both,
theory and simulations, coincide in that the plot of
$\tau^*_\alpha(t_w,\phi)$ as a function of $\phi$ for a given
fixed waiting time $t_w$ exhibits two regimes. The first
corresponds to samples that have fully equilibrated within this
waiting time $(\phi\le \phi_{c.o.}(t_w))$, and the second
corresponds to samples for which equilibration is not yet complete
$(\phi\ge \phi_{c.o.}(t_w))$. The rather loose boundary between
these two regimes defines a crossover volume fraction denoted by
$\phi_{c.o.}(t_w)$, illustrated by
the asterisks in Fig. \ref{fig3}, which increases with $t$ but
seems to saturate to the value $\phi_c \equiv
\phi_{c.o.}(t_w\to\infty) \approx 0.582$ determined by the
equilibrium SCGLE theory, as indicated in the
inset.

\section{Conclusions}\label{section4}

In summary, from the theoretical and simulated non-equilibrium
results compared in Figs. \ref{fig:1b} and \ref{fig3} we can
infer some important conclusions, but also identify several
equally relevant issues left open for further discussion. For
example, the comparison in Fig. \ref{fig:1b} confirms that, at
least within the window of waiting times considered, the NE-SCGLE
theory captures the correct kinetics of the simulated dynamic
arrest transition. This applies particularly to the characteristic
feature of the aging of glassy materials observed in the
non-equilibrium simulations, namely, the progressive development
with waiting time $t_w$, of the two-step decay of
$F_S(k,\tau;t_w)$ with correlation time $\tau$. Second, the
kinetic perspective provided by the non-equilibrium simulations
and by the NE-SCGLE theory, defines a useful additional conceptual
tool to describe some  aspects of the glass transition in model HS
liquids. For example, the ``equilibrated to non-equilibrated''
crossover in the $t_w$-dependence of $\tau^*_{\alpha}(t_w,\phi)$ in
Fig. \ref{fig3} could also be interpreted as a ``fragile to
strong'' dynamic crossover that changes with the age of the system
\cite{fscrossoverchen}. This opens the question of the relevance
of this NE-SCGLE scenario in the understanding of this actual
experimental ``fragile to strong'' dynamic crossover phenomena
observed in many molecular glass-formers \cite{fscrossoverchen}.
This discussion will be facilitated by the NE-SCGLE theory,
adapted here to polydisperse or multicomponent atomic liquids, but
that must still be extended to thermal protocols involving finite
cooling rates.

For the time being, however, the qualitative and quantitative
agreement in the comparison in Fig. \ref{fig2} illustrates the
overall consistency between the general scenario observed in the
simulations and that predicted by the NE-SCGLE theory. This
quantitative test, together with the qualitative consistency with
experimental observations of the recently-predicted NE-SCGLE
scenario of dynamically arrested spinodal decomposition, provide
encouraging evidences of the pertinence and accuracy of this
theoretical approach for the description of non-equilibrium
dynamic arrest phenomena.

\bigskip

$^{\star}$ Whitacre College of Engineering, Department of Chemical Engineering,
Texas Tech University, Lubbock, TX 79409-3121, USA.

$^{\dagger}$ Shull Wollan Center $-$ Joint Institute for Neutron Sciences, Oak
Ridge, TN 37831, USA.

$^{*}$ Departamento de Ingenier\'ia F\'isica, Divisi\'on de Ciencias
e Ingenier\'ias, Universidad de Guanajuato, Loma del Bosque 103, 37150 Le\'on, M\'exico.

\section*{Acknowledgments}
This work was supported by the Consejo Nacional de Ciencia y
Tecnolog\'{\i}a (CONACYT, M\'{e}xico) through grants  No. 242364,
182132, FC-2015-2/1155, Catedras CONACyT-1631,LANIMFE-279887-2017
and CB-2015-01-257636. P.M-M. acknowledges the Secretar\'{\i}a de
Educaci\'{o}n P\'{u}blica (SEP-PRODEP, M\'{e}xico) for a Post-doctoral
fellowship (DSA/103.5/15/1694 and DSA/105.5/16/4274) and
Postdoctoral fellowship from M\'exico Government (CONACYT, support 454743).
M.M.-N. acknowledges
the hospitality of Prof. Ram\'on Casta\~neda-Priego and the
support of the Universidad de Guanajuato (through the Convocatoria
Institucional para el Fortalecimiento de la Excelencia Acad\'emica
2015, project ``Statistical Thermodynamics of Matter Out
Equilibrium'').

\appendix

\section{Multicomponent extension of the NE-SCGLE theory (review of Ref.
\cite{nescgle4})}\label{appA}

The extension of the NE-SCGLE theory to multicomponent liquid was
developed in Ref. \cite{nescgle4}. The structural and dynamical
properties of a multicomponent liquid are written in terms of the
partial static structure factors $S_{\alpha \beta}(k;t_w)$ and
partial (collective and self) intermediate scattering functions,
$F_{\alpha \beta}(k,\tau;t_w)$ and
$F^S_{\alpha\beta}(k,\tau;t_w)$, of the binary mixture. The
determination of these partial properties involve the solution of
the multicomponent version \cite{nescgle4} of Eqs.
(\ref{relsigmadif2pp})- (\ref{lambdadk}) above.

For an $s$-component mixture, these equations are
\begin{eqnarray}
\frac{\partial S(k;t_w)} {\partial t_w} = -k^2 D^0 \cdot b(t_w)
\cdot
\Bigl[\sqrt{n} \cdot \mathcal{E}(k;n,T_f) \cdot  \nonumber\\
\sqrt{n}\Bigr] \cdot S(k;t_w) - S(k;t_w) \cdot
\Bigl[\sqrt{n} \cdot \mathcal{E}(k;n,T_f) \cdot  \nonumber\\
\sqrt{n}\Bigr] \cdot b(t_w) \cdot D^0 k^2 +2k^2 D^0 \cdot b(t_w),
\label{evolskt}
\end{eqnarray}
with $\sqrt{n}$ being a $s\times s$ diagonal matrix whose
$\alpha$th diagonal element is  $\sqrt{n_{\alpha}}$, and in which
the element $\mathcal{E}_{\alpha\beta}(k;n,T_f)$ of the matrix
$\mathcal{E}(k;n,T_f)$ is the Fourier transform (FT) of the
functional derivative
$\mathcal{E}_{\alpha\beta}[\textbf{r}-\textbf{r}';n,T] \equiv
\left[ {\delta \beta\mu_\alpha [{\bf r};n,T]}/{\delta n_\beta({\bf
r}')}\right] \ = \delta({\bf r}-{\bf r}')/ n_{\alpha}({\bf r})
-c_{\alpha\beta}^{(2)}[{\bf r}, {\bf r}';n,T]$ evaluated at the
(fixed) composition $n=(n_1,...,n_s)$ and final temperature $T_f$
of the quenched system, with $c_{\alpha\beta}^{(2)}[{\bf r}, {\bf
r}';n,T]$ being  the \emph{direct correlation function}. The
non-zero elements of the $s \times s$ diagonal matrices $D^0$ and
$b(t_w)$ are, respectively, the short-time self-diffusion
coefficients $D^0_\alpha$ and time-dependent mobility functions
$b_\alpha(t_w)$,  of species $\alpha$. The latter is written as
\begin{equation}
b_\alpha[\tau; t_w]= \left[1+\int_0^\infty  d\tau
\Delta\zeta^*_\alpha[\tau; t_w]\right]^{-1}, \label{bast5p}
\end{equation}
with  $\Delta\zeta^*_\alpha[\tau; t_w]$ approximated by
\begin{eqnarray}
\Delta \zeta^*_{\alpha} (\tau; t_w)=\frac{D^0_\alpha}{3\big( 2\pi
\big) ^{3}}\int d {\bf k}\ k^2 [F^S(\tau)]_{\alpha\alpha}\Bigl[h
\cdot  \sqrt{n}  \cdot  \nonumber\\ S^{-1} \cdot F(\tau) \cdot
S^{-1} \cdot   \sqrt{n}\cdot h \Bigr]_{\alpha\alpha}.
\label{dzdtppp}
\end{eqnarray}
In this equation the matrix $h$ is given by $ h
=\sqrt{\overline{n}}^{-1} \cdot (S-I) \cdot
\sqrt{\overline{n}}^{-1}$ and we have systematically omitted the
arguments $k$ and $t_w$ of the $s\times s$ matrices $h(k;t_w)$,
$S(k;t_w)$, $F(k,\tau;t_w)$, and $F^S(k,\tau;t_w)$. Finally, the
time-evolution equations for $F(k,\tau;t_w)$ and $F^S(k,\tau;t_w)$
in Laplace space read
\begin{eqnarray}\label{fluct5ppp}
&&\hat{F}(k,z; t_w) = \lbrace zI + k^2 D^0 \cdot [ zI + \nonumber \\
&& \lambda(k;t_w) \cdot  \Delta \hat{\zeta}^{*}(z; t_w) ]^{-1}
\cdot
 S^{-1}(k;t_w) \rbrace^{-1} \cdot S(k; t_w)  \nonumber \\
\end{eqnarray}
and
\begin{eqnarray} \label{fluct5sppp}
&& \hat{F^S(k,z; t_w)} = \lbrace zI + k^2 D^0 \cdot [ zI + \nonumber \\
&& \lambda(k;t_w)  \cdot \Delta \hat{\zeta}^{*}(z; t_w)]^{-1} \rbrace^{-1} \nonumber \\
\end{eqnarray}
where $\hat F(k,z; t_w)$ and $\hat F^S(k,z; t_w)$ are  the Laplace
transforms of the collective and self partial intermediate
scattering functions $F_{\alpha\beta}(k,\tau; t_w)$ and
$F_{\alpha\beta}^S(k,\tau; t_w)$, and $\lambda (k;t_w)$ is a
diagonal matrix whose  non-zero elements $\lambda _{\alpha
\alpha}(k;t_w)$ are given by

\begin{gather}
\label{fluct6sppp} \lambda _{\alpha
\alpha}(k;t_w)=1/\left[1+\left( k/k^{c}_\alpha(t_w)\right)
^{2}\right].
\end{gather}
Eqs. (\ref{evolskt})- (\ref{fluct6sppp}) constitutes the essence
of the non-equilibrium self-consistent generalized Langevin
equation  (NE-SCGLE) theory describing the irreversible  isochoric
relaxation of a suddenly quenched liquid mixture with underlying
Brownian or diffusive short-time microscopic dynamics.

\section{Molecular adaptation (following Ref.
\cite{AtomicMix}).}\label{appB}

The theoretical predictions presented and discussed in the present
paper involve one additional correction, namely, the introduction
of a simple interpolating device to incorporate the correct
short-time ballistic  limit of the dynamics of atomic liquids in
the NE-SCGLE dynamic properties (illustrated in Fig. \ref{fig:1b}). This
correction does not affect the essential features of the predicted
long-time dynamics associated with the glass transition. However,
it is needed to compare the theory, developed for Brownian liquids
with underlying short-time  diffusive microscopic dynamics, with
the results of \emph{molecular} dynamics simulations, whose
short-time dynamics is ballistic. This issue is thus not inherent
to the non-equilibrium nature of the NE-SCGLE theory, and in fact,
it has recently been discussed in more detail  in Ref.
\cite{AtomicMix} in the context of the \emph{equilibrium} SCGLE
theory. In the present work we assume that exactly the same
arguments and approximations apply when adapting the NE-SCGLE
theory of multicomponent Brownian liquids, summarized in the
previous section (Eqs. (\ref{evolskt})-(\ref{fluct6sppp})), to the
description of the dynamics of multicomponent  \emph{atomic}
liquids.

In essence, following Ref. \cite{AtomicMix}, we use the fact that
the NE-SCGLE equations (Eqs. (\ref{evolskt})-(\ref{fluct6sppp}))
also describe the non-equilibrium dynamics of the atomic mixture
in the long-time diffusive regime, and that a simple manner to
interpolate between the correct short-time ballistic and long-time
difusive behavior, is provided by the interpolating expressions in
Eqs. (4.4)-(4.6)  of Ref. \cite{AtomicMix}. In the present
non-equilibrium context, the first of these equations is an
integro-diferential equation for the mean square displacement
$W_\alpha^{(molec)}(\tau;t_w)$,
\begin{eqnarray}
\frac{M_\alpha}{\zeta^0_\alpha}\frac{dW_\alpha^{(molec)}(\tau;t_w)}{d\tau}
+W_\alpha^{(molec)}(\tau;t_w) =D^0_\alpha\tau- \nonumber\\
\int_0^\tau \Delta \zeta^*_\alpha (\tau-\tau';t_w)
W_\alpha^{(molec)}(\tau';t_w)d\tau', \label{wdtintdifec}
\end{eqnarray}
where $M_\alpha$ is the mass and $\zeta^0_\alpha=k_BT/D_\alpha^0$,
with $D_\alpha^0$ being the short-time self-diffusion coefficient
of the $\alpha$th  atomic species and $T$ being the final
temperature of the quench.

The solution of this equation for $W_\alpha^{(molec)}(\tau;t_w)$
satisfies the correct short-time ballistic limit. Introduced in
the format of a Gaussian approximation, it guarantees the correct
short-time ballistic limit of the collective and self ISFs. To use
this fact we follow Eqs. (4.5) and (4.6)  of Ref.
\cite{AtomicMix}, which in our non-equilibrium context   are
written as the following approximate interpolating expressions for
the $s\times s$ matrices $F^{(molec)}(k,\tau;t_w)$ and
$F_S^{(molec)}(k,\tau;t_w)$.
\begin{align}
F^{(molec)}(k,\tau;t_w)= & F(k,\tau;t_w)+\nonumber \\
& \{ S(k,t_w) \cdot \exp [-k^2W^{(molec)}(\tau;t_w)\cdot  \nonumber\\
& S^{-1}(k,t_w)] -F(k,\tau;t_w)\}\cdot \exp[-Z
\tau],\label{fktinterpolation}
\end{align}
and
\begin{align}
F_S^{(molec)}(k,\tau;t_w)=& F_S(k,\tau;t_w)+\{\exp [-k^2W^{(molec)}(\tau;t_w)]\nonumber\\
& -F_S(k,\tau;t_w)\}\cdot \exp[-Z \tau]. \label{fsktinterpolation}
\end{align}
In these ($s \times s$) \emph{matrix } equations,  the diagonal
matrices $W^{(molec)}(\tau;t_w)$ and $Z$ have diagonal elements
$W^{(molec)}_{\alpha }(\tau;t_w)$ and $Z_{\alpha } \equiv
(\zeta^0_\alpha/M_\alpha) $, respectively.

The resulting \emph{molecular} version of the multicomponent
NE-SCGLE theory is thus contained in Eqs.
(\ref{evolskt})-(\ref{fluct6sppp}) plus Eqs.
(\ref{wdtintdifec})-(\ref{fsktinterpolation}). The solution of
these equations provides a first-principles description of the
main dynamic properties of a simple molecular liquid mixture. In a
specific application, we start by solving Eqs.
(\ref{evolskt})-(\ref{fluct6sppp}) to determine $\Delta
\zeta^*(\tau;t_w)$, $F(k,\tau;t_w)$, and $F_S(k,\tau;t_w)$. These
functions describe the short-$\tau$ \emph{diffusive} dynamics of
Brownian, not molecular liquids. To incorporate the correct
short-time ballistic limit, we employ these functions as input of
Eqs. (\ref{wdtintdifec})-(\ref{fsktinterpolation}), thus
evaluating $F^{(molec)}(k,\tau;t_w)$, $F_S^{(molec)}(k,\tau;t_w)$,
and $W_\alpha^{(molec)}(\tau;t_w)$. These functions describe the
predicted NE-SCGLE dynamics of our atomic or molecular mixture.
There, however, we have omitted the superscript $(molec)$, only
employed here for the clarity of the present summary.

\section{Modeling polydispersity: the atomic hard-sphere
liquid.}\label{appC}

In order to actually practice the protocol outlined in the last
paragraph to solve the NE-SCGLE Eqs.
(\ref{evolskt})-(\ref{fsktinterpolation}), there are still a few
elements that await a more accurate definition. We refer to the
short-time self-diffusion coefficients $D_\alpha^0$, to the cutoff
wave-vectors $k^{c}_\alpha(t_w)$ entering in the interpolating
functions in Eq. (\ref{fluct6sppp}), and to the matrix
$\mathcal{E}(k;n,T_f)$. These elements, however, are
system-dependent, and hence, must be determined in the context of
the concrete model system studied. Thus, let us now address this
issue in the context of the monocomponent (but polydisperse)
hard-sphere liquid discussed in the paper. This system is modeled
in the simulations as a monocomponent but polydisperse hard-sphere
liquid with HS diameters subjected to a continuous uniform
distribution yielding a polydispersity of 8.66 \%.

In the theoretical modeling we approximate this uniform
distribution by a binodal distribution yielding the same
polydispersity, i.e., as an equimolar binary HS mixture with
diameters $\sigma_1= (1 - \epsilon)$  and  $ \sigma_2 = (1 +
\epsilon)$, with $\epsilon=0.0866$. Hence, the structural and
dynamic properties of the resulting bidisperse liquid,
$S(k;t_w)=\sum_{\alpha,\beta=1}^2 \sqrt{x_\alpha x_\beta}
S_{\alpha \beta}(k;t_w)$, $F(k,\tau;t_w)=\sum_{\alpha, \beta=1}^2
\sqrt{x_\alpha x_\beta}F_{\alpha \beta}(k,\tau;t_w)$, and
$F_S(k,\tau;t_w) = \ \sum_{\alpha=1}^2 x_\alpha F^S_{\alpha
}(k,\tau;t_w)$, are written in terms of the partial static
structure factors $S_{\alpha \beta}(k;t_w)$ and partial
(collective and self) intermediate scattering functions,
$F_{\alpha \beta}(k,\tau;t_w)$ and $F^S_{\alpha
\beta}(k,\tau;t_w)$ of the binary mixture.

The determination of $D_\alpha^0$, $k^{c}_\alpha(t_w)$, and
$\mathcal{E}(k;n,T_f)$ must be made at the level of the
\emph{equilibrium} version of the theory. For this we mean the
long-$t_w$ asymptotic limit of Eqs.
(\ref{evolskt})-(\ref{fsktinterpolation}) in which the matrix
$S(k;t_w)$ has reached the \emph{equilibrium} stationary solution
of Eq. (\ref{evolskt}), namely, $S(k;t_w\to \infty) \equiv
S^{eq}(k;n,T_f) = \left[ \sqrt{n} \cdot \mathcal{E}(k;n,T_f) \cdot
\sqrt{n} \right]^{-1}$. In this limit,  Eqs.
(\ref{dzdtppp})-(\ref{fsktinterpolation}) become a closed system
of equations for the equilibrium dynamic properties
$F^{eq}(k,\tau)$, $F_S^{eq}(k,\tau)$, and $W_\alpha^{eq}(\tau)$,
given $S^{eq}(k;n,T)$ as input. This equilibrium theory was
developed in Ref. \cite{AtomicMix} and applied there to the
prediction of the equilibrium properties of the same
\emph{polydisperse} hard-sphere liquid discussed in this work. For
this, the assumption was made that
\begin{equation}
D^0_1\approx D^0_2\approx D^0\equiv \frac{3}{8
}\left(\frac{k_BT}{\pi M}\right)^{1/2}\frac{1}{n\bar{\sigma}^2},
\label{dkinetictheory}
\end{equation}
and the equilibrium partial static structure factors
$S^{eq}_{\alpha \beta}(k)$ were provided by their
Percus-Yevick-Verlet-Weis (PYVW) approximation
\cite{percusyevick,verletweis}, adapted to multicomponent fluids
in Ref. \cite{williams}. Then the cutoff wave-vectors
$k^{c}_\alpha$ were written as $k^{c}_\alpha = 1.119
k^{max}_\alpha$, with $k^{max}_\alpha$ being the position of the
main peak of $S^{eq}_{\alpha \alpha}(k)$.

Going back to the full non-equilibrium theory employed in this
work, in the NE-SCGLE Eqs.
(\ref{evolskt})-(\ref{fsktinterpolation}), we adopt the same
equilibrium definition of $D_\alpha^0$ in Eq.
(\ref{dkinetictheory}), whereas the matrix $\mathcal{E}(k;n,T_f)$
needed as input in these equations is determined by the
equilibrium condition $\mathcal{E}(k;n,T_f)= \left[ \sqrt{n} \cdot
S^{eq}(k;n,T_f) \cdot  \sqrt{n}\right]^{-1}$, with
$S^{eq}(k;n,T_f)$ also approximated by its multicomponent
Percus-Yevick-Verlet-Weis (PYVW) approximation
\cite{percusyevick,verletweis,williams}. As for the cutoff wave
vector $k_c(t_w)$, we also adopt the equilibrium prescription, so
that  $k_c(t_w)=1.119 \times k^{max}_\alpha(t_w)$, with
$k^{max}(t_w)$ being the the position  of the main peak of
$S(k,t_w)$.

\section{Non-equilibrium molecular dynamics simulations.} \label{appD}
In this work we performed non-equilibrium molecular dynamics
(NE-MD) simulations to describe the non-equilibrium structural and
dynamical evolution of a polydisperse hard-sphere system in their
metastable regime close to the glass transition. Our NE-MD
simulation data are produced using event-driven simulations and
following the same methodology explained in Ref. \cite{gabriel}.
We have used polydisperse samples whose diameters are evenly
distributed between $\overline{ \sigma} (1-w/2)$ and $\overline{
\sigma} (1+w/2)$, with $\overline \sigma$ being the mean diameter.
In this study, as in the previous work, we have considered the
case $w=0.3$, corresponding to a polydispersity $s_\sigma =
w/\sqrt{12}=0.0866$. The initial configurations are prepared by
placing $N$-soft spheres at completely random positions in a cubic
cell of volume $V$,  interacting through a short ranged repulsive
soft (but increasingly harder) interaction and in the presence of
strong dissipation, and all the particles are assumed to have the
same mass $M$. These nonthermalized hard-sphere configurations are
then given random velocities taken from a Maxwell-Boltzmann
distribution, with $k_BT$ set as the energy unit, and are used as
the starting configurations for the event-driven simulations. All
results are showed in reduced units, i.e., lenght in units of
$\overline{\sigma}$, time in units of
$\overline{\sigma}\sqrt{M/k_BT}$.

With the purpose of completing our study about the equilibration
and aging of a \emph{polydisperse} hard-sphere (HS) liquid, which
is described in Ref. \cite{gabriel}, we investigate the
finite-size effect on the non-equilibrium structural and dynamical
evolution of the polydisperse system. We have run simulations over
systems of $N$=1024, 2048, 4096 and 8192 spheres and with the
intention to generate a reasonable statistical, we have run at
least $40$ independent realizations for an array of volume
fractions between $0.55$ and $0.58$.
\medskip

\begin{figure}[h]
\centering
\subfigure{ \label{fig:4a}
\includegraphics[scale=.25]{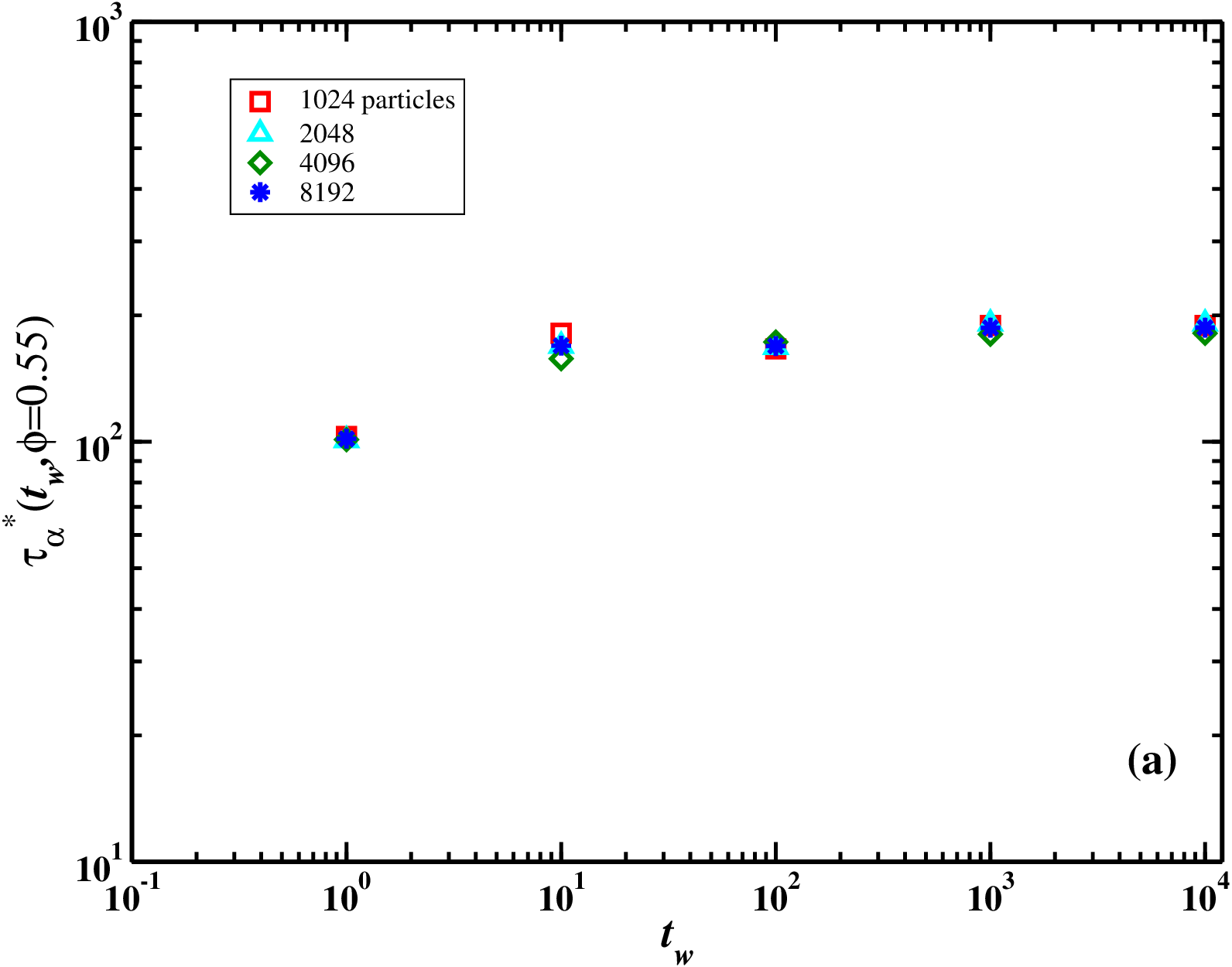}
}
\centering
\subfigure{ \label{fig:4b}
\includegraphics[scale=.25]{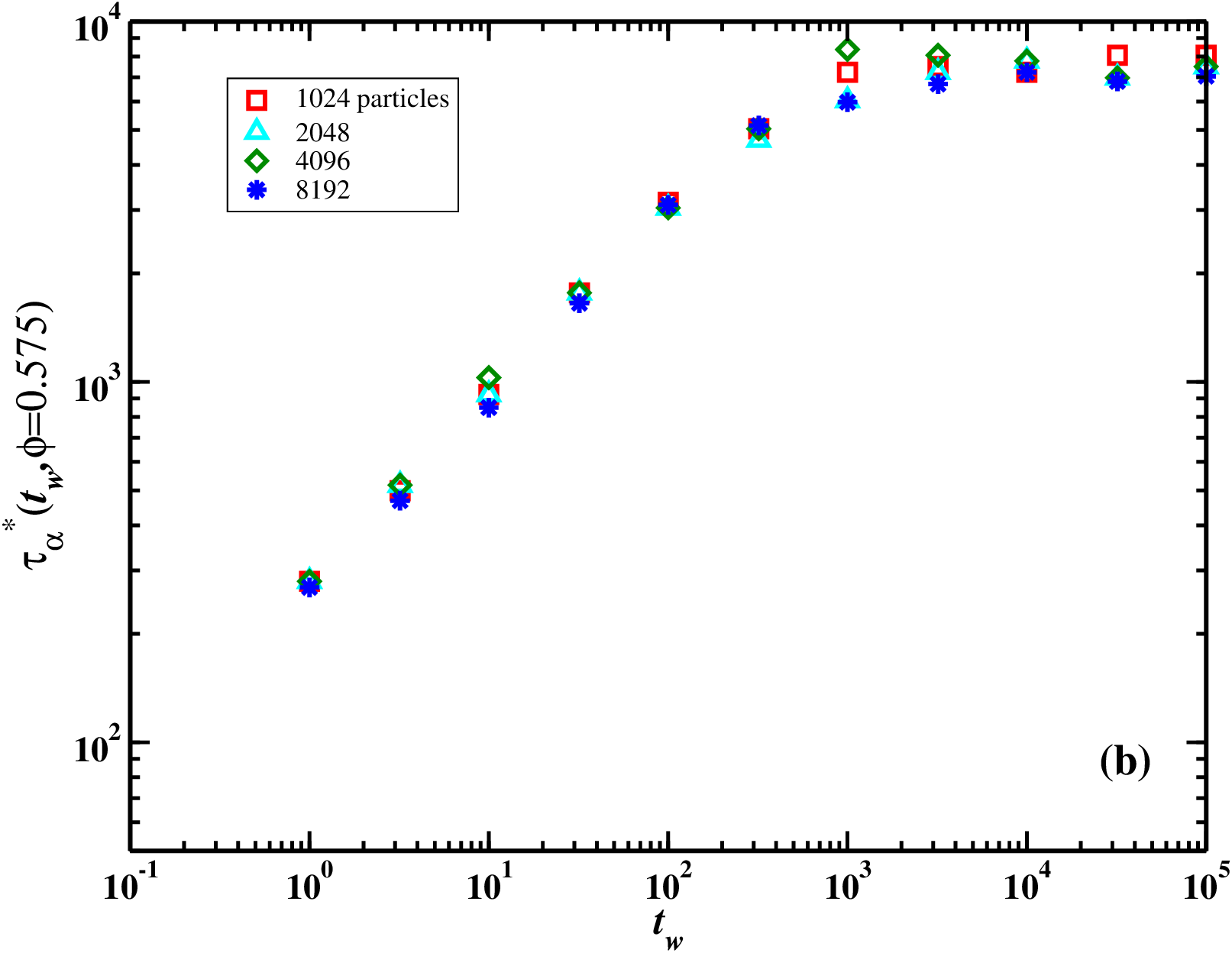}
}
\caption{Nonequilibrium evolution of the dimensionless
$\alpha$-relaxation time, displayed as
$\tau_{\alpha}^*(k=7.1;t_w)$ corresponding to the equilibration
processes at fixed volume fractions Fig. \ref{fig:4a} $\phi=0.55$ and Fig. \ref{fig:4b}
$\phi=0.575$.} \label{fig:1SM}
\end{figure}

Our main conclusions are as follow: First, the results obtained do
not show a significant dependence on particle number, at least not
in all metastable regime, and are independent of the number of
realizations. Second, the results are consistent with those
reported in Ref. \cite{gabriel}.  This is further illustrated in
the Fig. \ref{fig:1SM} in which we plot the $\alpha$-relaxation
time $\tau^*_{\alpha}(t_w,\phi)$, defined by the condition
$F_s(k,\tau_{\alpha};t_w)=e^{-1}$, as a function of the evolution
time $t_w$ for two distinct, representive volume fractions of the
metastable regime $\phi$=0.55 and 0.575. As can be noted on the
figure, in the case of $\phi=0.55$, the data almost overlap each
other for the waiting times considered. In the case of
$\phi=0.575$, a slight difference can be observed for higher times
than $t_w=10^3$.

Due to the enormous amount of time required to run the simulation
for volume fractions $\phi>0.58$, we have decided not to include
the preliminary results here but suggest that the loss of
ergodicity becomes a truly fundamental challenge, since the size
of the representative non-equilibrium ensemble needed to get
stable statistics in the simulations seems to increase without
bound as one gets deeper in the glassy regime. This is indeed work
in progress, but we believe that the discussion of the paper does
exhibit an immediate contribution of the \emph{non-equilibrium}
SCGLE theory, namely, the conceptual enrichment of the discussion
of the glass transition problem by introducing the waiting-time
dimension $t_w$ in the description of glassy behavior.

\end {document}